\documentclass[letterpaper]{jpconf}
\usepackage{graphicx}
\begin{document}
\title{Ab Initio Theory of Light-ion Reactions}

\author{Petr Navr\'atil$^1$, Sofia Quaglioni$^1$, and Robert Roth$^2$}

\address{$^1$Lawrence Livermore National Laboratory, P.O. Box 808, L-414, Livermore, CA 94551, USA}

\address{$^2$Institut f\"{u}r Kernphysik, Technische Universit\"{a}t Darmstadt, 64289 Darmstadt, Germany}

\ead{navratil1@llnl.gov}

\begin{abstract}
The exact treatment of nuclei starting from the constituent nucleons and the fundamental interactions among them has been a long-standing goal in nuclear physics. Above all nuclear scattering and reactions, which require the solution of the many-body quantum-mechanical problem in the continuum, represent a theoretical and computational challenge for {\it ab initio} approaches. After a brief overview of the field, we present a new {\it ab initio} many-body approach capable of describing simultaneously both bound and scattering states in light nuclei. By combining the resonating-group method (RGM) with the {\it ab initio} no-core shell model (NCSM), we complement a microscopic cluster technique with the use of realistic interactions and a microscopic and consistent description of the clusters. We show results for neutron and proton scattering on light nuclei, including $p$-$^7$Be and $n$-$^8$He. We also highlight the first results of the  $d$-$^3$He and $d$-$^3$H fusion calculations obtained within this approach.
\end{abstract}

\section{Introduction}

Nuclei are quantum many-body systems with both bound and unbound states. A realistic {\it ab initio} approach to light nuclei with predictive power must have the capability to describe both classes of states within a unified framework. Over the past 15 years, significant progress has been made in our understanding of the properties of the bound states of light nuclei starting from realistic nucleon-nucleon ($NN$) interactions, see e.g. Ref.~\cite{benchmark} and references therein, and more recently also from $NN$ plus three-nucleon ($NNN$) interactions~\cite{Nogga00,GFMC1,GFMC2,GFMC3,NO03}. However, the calculation of unbound states and nuclear reactions represents a formidable challenge for nuclear theory, the main obstacle being the treatment of the scattering states. 

A predictive theory of reactions of light nuclei is needed for many reasons. First, it would greatly help our understanding of nuclear reactions important for astrophysics. As an example, the $^7$Be($p$,$\gamma$)$^8$B radiative capture is a step not known accurately enough in the nucleosynthetic chain leading to $^8$B, which in turn is the dominant source of the high-energy solar neutrinos (through $\beta$-decay to $^8$Be) detected in terrestrial experiments. An even more dramatic example is provided by the $^8$Be($\alpha$,$\gamma$)$^{12}$C and $^{12}$C($\alpha$,$\gamma$)$^{16}$O radiative capture reactions. The ratio of the thermonuclear reaction yields for these two processes determines the carbon-to-oxygen ratio at the end of helium burning with important consequences for the production of all species made in subsequent burning stages in the stars. At stellar energies ($\sim$300 keV) radiative capture rates are too small to be measured in the laboratory.  Thus, measurements are performed at higher energies and extrapolations to the energy region of interest using theory are unavoidable. The theoretical extrapolation is, however, challenging due to the influence of several resonances. A fundamental theory would be of great use here. 

Furthermore, nuclear reactions are one of the best tools for studying exotic nuclei, which have become the focus of the next generation experiments with rare-isotope beams. These are nuclei for which most low-lying states are unbound, so that a rigorous analysis requires scattering boundary conditions. In addition, much of the information we have on the structure of these short-lived systems is inferred from reactions with other nuclei.

Finally,  low-energy fusion reactions represent the primary energy-generation mechanism in stars, and could potentially be used for future energy generation on earth. Examples of these latter reactions include the $d$+$^3$H$\rightarrow n$+$^4$He fusion used at ITER and at the National Ignition Facility (NIF). Even though there have been many experimental investigations of the cross sections of this reaction, there are still open issues. A first-principles theory will provide the predictive power to  reduce the uncertainty in the reaction rate at very low temperatures; provide an understanding of the reaction rate dependence on the polarization induced by the strong magnetic fields (characteristic of both inertial and magnetic confinement); and clarify the influence of non-local thermal equilibrium in plasma environments. 

In this contribution, we present a brief overview of the field in Sect.~\ref{overview}. In Sect.~\ref{NCSM_RGM}, we describe the recently introduced  {\it ab initio} many-body approach to reactions on light nuclei that combines  the resonating-group method (RGM)~\cite{RGM} with the {\it ab initio} no-core shell model (NCSM)~\cite{NCSM}. In Sect.~\ref{results}, after presenting our results for nucleon-$^4$He and $n$-$^3$H scattering, we discuss examples of calculations relevant for nuclear astrophysics, $p$-$^7$Be scattering, investigate an exotic nucleus, $^9$He, and, finally, highlight our first calculations of the  $d$-$^3$He and $d$-$^3$H fusion reactions relevant for energy generation.  Conclusions are given in Sect.~\ref{concl}.

\section{Microscopic approaches to nuclear reactions}\label{overview}

Because of their importance nuclear reactions attract much attention, and there have been many interesting new developments in the recent past. In this section we will give a brief and non-exhaustive overview of the theoretical efforts devoted to microscopic approaches to nuclear reactions, and in particular scattering of light nuclei. 

By microscopic approaches we mean methods, in which all the nucleons involved in the process are treated as active degrees of freedom, and the antisymmetrization of the many-body wave functions is treated exactly. If the $NN$ interactions among all participating nucleons are realistic, i.e., describe accurately $NN$ scattering and the deuteron, we call the calculation {\it ab initio}.  

In the three- and four-nucleon sectors there has been remarkable progress over the past decade: the Faddeev~\cite{Witala01}, Faddeev-Yakubovski (FY)~\cite{Lazauskas04,Lazauskas09}, Alt-Grassberger and Sandhas (AGS)~\cite{Deltuva_a,Deltuva_b}, hyperspherical harmonics (HH)~\cite{Marcucci09}, Lorentz integral transform methods~\cite{Gazit,chiral_LIT_4He,Bacca09}, RGM~\cite{Hofmann08}, etc., are among the best known of several numerically exact techniques able to describe reactions observables starting from realistic $NN$ and in some cases also $NNN$ forces. 

Going beyond four nucleons there are fewer {\it ab initio} or {\it ab initio} inspired methods able to describe reactions observables starting from realistic forces. Only very recently the Green's function Monte Carlo (GFMC)~\cite{GFMC_nHe4}, the no-core shell model combined with the resonating group method (NCSM/RGM)~\cite{NCSM_RGM,NCSM_RGM_long} and the fermionic molecular dynamics~\cite{FMD} have made steps in this direction.

Reactions among light nuclei are more widely described starting from semi-realistic $NN$ interactions with adjusted parameters within the traditional resonating-group method~\cite{RGM1,RGM2,RGM3} or the generator coordinate method~\cite{Desc04,Baye05,Damman09}, which are microscopic cluster techniques. In these approaches, the $NN$ interactions are often simplified to central forces and the cluster wave functions are not necessarily eigenstates obtained with the same interaction as that used between nucleons from different clusters. 

Among the recent developments in the $A=4$ scattering and reaction calculations we highlight the new capability to include properly the Coulomb interaction in momentum space~\cite{Deltuva_a,Deltuva_b} and to include the three-nucleon interaction in the $p$-$^3$H Faddeev-Yakubovski configuration space calculations~\cite{Lazauskas09}. A benchmark for the $n$-$^3$H low-energy elastic cross section calculation has been performed by the FY, AGS and HH methods using different $NN$ potentials~\cite{Lazauskas05}. The main conclusion of this work is the failure of the existing $NN$ forces to reproduce the $n$-$^3$H total cross section. Remarkable recent results are the $p$-$^3$He scattering calculations performed using the hyperspherical harmonic basis, which demonstrated that the new $NN$ plus $NNN$ interactions derived within chiral effective field theory (EFT) reduce noticeably the discrepancy observed for the $A_y$ observable~\cite{Viviani10}. Further, with the same Hamiltonian, the low-energy total $n$-$^3$H cross section calculated by the HH method was found in improved agreement with the data~\cite{Viviani09}.

The first {\it ab initio} scattering calculation for a system with $A>4$ was performed within the GFMC approach. The $n$-$\alpha$ low-lying  $J^\pi=3/2^-$ and $1/2^-$ $P$-wave resonances as well as the $1/2^+$ $S$-wave non-resonant scattering below 5 MeV center of mass (c.m.)\ energy were obtained using the AV18 $NN$ potential with and without the three-nucleon force, chosen to be either the Urbana IX or the Illinois-2 model~\cite{GFMC_nHe4}.  The results of these calculations revealed sensitivity to the inter-nucleon interaction, and in particular to the strength of the spin-orbit force.

As an interesting theoretical development to the many-body scattering, we mention the approach based on the variational description of continuum states in terms of integral relations~\cite{Kievsky10} that may be used to directly apply the bound-state many-body techniques to scattering. A variation of this approach has been explored in the $A=5$ scattering in Ref.~\cite{Suzuki09}.

There are also some recent attempts to describe the nuclear scattering in an effective field theory approach. In particular, the pion-less EFT combined with the RGM was successfully applied to three- and four-nucleon bound state and scattering calculations~\cite{Kirscher10}.

In a big jump in mass number, we note that the $^{17}$F low-lying states were recently investigated within the coupled-cluster approach with the Gamow-Hartree-Fock basis that incorporates effects of the continuum~\cite{HPH10}. The calculation resulted in a good description of the $1/2^+$ proton halo state in $^{17}$F. It was shown that the continuum effects are essential to obtain these results and that the proton halo state single-particle energy is not affected by short-range correlations in the nuclear interactions. 

\section{Ab initio NCSM/RGM}\label{NCSM_RGM}

We have developed a new many-body approach, the {\it ab initio} NCSM/RGM~\cite{NCSM_RGM,NCSM_RGM_long}, capable of simultaneously describing both bound and scattering states in light nuclei throughout the $p$-shell, thus, filling the gap between the $A=4$ and the $A=17$ systems. The NCSM~\cite{NCSM} is an approach to the microscopic calculation of ground and low-lying excited states of light nuclei with realistic two- and, in general, three-nucleon forces.  The RGM~\cite{RGM} is a microscopic cluster technique based on the use of $A$-nucleon Hamiltonians, with fully anti-symmetric many-body wave functions built assuming that the nucleons are grouped into clusters. Although most of its applications are based on the use of binary-cluster wave functions, the RGM can be formulated for three (and, in principle, even more) clusters. The use of the harmonic oscillator (HO) basis in the NCSM results in an incorrect description of the wave-function asymptotic and a lack of coupling to the continuum.  By combining the NCSM with the RGM, we complement the ability of the RGM to deal with scattering and reactions with the utilization of realistic interactions and a consistent microscopic description of the nucleonic clusters achieved via {\it ab initio} NCSM, while preserving important symmetries such as Pauli exclusion principle, translational invariance, and angular momentum.

Here, we limit ourselves to a two-cluster RGM, which is based on binary-cluster channel states of total angular momentum $J$, parity $\pi$, and isospin $T$,
\begin{equation}
|\Phi^{J^\pi T}_{\nu r}\rangle = \Big [ \big ( \left|A{-}a\, \alpha_1 I_1^{\,\pi_1} T_1\right\rangle \left |a\,\alpha_2 I_2^{\,\pi_2} T_2\right\rangle\big ) ^{(s T)}
\,Y_{\ell}\left(\hat r_{A-a,a}\right)\Big ]^{(J^\pi T)}\,\frac{\delta(r-r_{A-a,a})}{rr_{A-a,a}}\,.\label{basis}
\end{equation}
In the above expression, $\left|A{-}a\, \alpha_1 I_1^{\,\pi_1} T_1\right\rangle$ and $\left |a\,\alpha_2 I_2^{\,\pi_2} T_2\right\rangle$ are the internal (antisymmetric) wave functions of the first and second cluster, containing $A{-}a$ and $a$ nucleons ($a{<}A$), respectively. They are characterized by angular momentum quantum numbers $I_1$ and $I_2$ coupled together to form the channel spin $s$. The relative coordinate $\vec r_{A-a,a}$ represents the separation of the cluster centers of mass. The channel states~(\ref{basis}) have relative angular momentum $\ell$. It is convenient to group all relevant quantum numbers into a cumulative index $\nu=\{A{-}a\,\alpha_1I_1^{\,\pi_1} T_1;\, a\, \alpha_2 I_2^{\,\pi_2} T_2;\, s\ell\}$.
 
The former basis states can be used to expand the many-body wave function according to
\begin{equation}
|\Psi^{J^\pi T}\rangle = \sum_{\nu} \int dr \,r^2\frac{g^{J^\pi T}_\nu(r)}{r}\,\hat{\mathcal A}_{\nu}\,|\Phi^{J^\pi T}_{\nu r}\rangle\,. \label{trial}
\end{equation}
As the basis states~(\ref{basis}) are not anti-symmetric under exchange of nucleons belonging to different clusters, in order to preserve the Pauli principle one has to introduce the appropriate inter-cluster antisymmetrizer $\hat{\mathcal A}_{\nu}$. The coefficients of the expansion~(\ref{trial}) are the relative-motion wave functions $g^{J^\pi T}_\nu(r)$, which represent the only unknowns of the problem. To determine them one has to solve the non-local integro-differential coupled-channel equations 
\begin{equation}
\sum_{\nu}\int dr \,r^2\left[{\mathcal H}^{J^\pi T}_{\nu^\prime\nu}(r^\prime, r)-E\,{\mathcal N}^{J^\pi T}_{\nu^\prime\nu}(r^\prime,r)\right] \frac{g^{J^\pi T}_\nu(r)}{r} = 0\,,\label{RGMeq}
\end{equation}
where the two integration kernels, the Hamiltonian kernel,
\begin{equation}
{\mathcal H}^{J^\pi T}_{\nu^\prime\nu}(r^\prime, r) = \left\langle\Phi^{J^\pi T}_{\nu^\prime r^\prime}\right|\hat{\mathcal A}_{\nu^\prime}H\hat{\mathcal A}_{\nu}\left|\Phi^{J^\pi T}_{\nu r}\right\rangle\,,\label{H-kernel}
\end {equation}
and the norm kernel,
\begin{equation}
{\mathcal N}^{J^\pi T}_{\nu^\prime\nu}(r^\prime, r) = \left\langle\Phi^{J^\pi T}_{\nu^\prime r^\prime}\right|\hat{\mathcal A}_{\nu^\prime}\hat{\mathcal A}_{\nu}\left|\Phi^{J^\pi T}_{\nu r}\right\rangle\,,\label{N-kernel}
\end{equation}
contain all the nuclear structure and anti-symmetrization properties of the problem. In particular, the non-locality of the kernels is a direct consequence of the exchanges of nucleons between the clusters. We have used $E$ and $H$ to denote the total energy in the center-of-mass frame and the intrinsic $A$-nucleon microscopic Hamiltonian, respectively.
  
The formalism presented above is combined with the {\em ab initio} NCSM in two steps. First, the cluster eigenstates appearing in Eq.~(\ref{basis}) are obtained by means of NCSM calculations with the intrinsic part of the realistic Hamiltonian $H$ used in Eq.~(\ref{H-kernel}). An HO basis truncated by the maximum number of the HO excitations $N_{\rm max}$ above the unperturbed ground state is used. 

Second, we replace the delta functions in the localized parts of the Hamiltonian~(\ref{H-kernel}) and the norm~(\ref{N-kernel}) kernels with their representation in the HO model space. We use the same HO frequency ($\Omega$) as for the cluster eigenstate wave functions and a consistent model-space size ($N_{\rm max}$). We emphasize that this replacement is performed only for the localized parts of the kernels. The diagonal parts coming from the identity operator in the antisymmetrizers, the kinetic term and the average Coulomb potential are treated exactly.

An advantage of the HO basis is its flexibility. In particular, if the $N_{\rm max}\hbar\Omega$ truncation is used, one may utilize the Slater-determinant (SD) single-particle HO basis and still preserve the translational invariance (for a translationally invariant Hamiltonian or other operator). One can even combine the SD basis and the relative-coordinate (Jacobi) basis. This is what we do in the NCSM/RGM applications to $p$-shell nuclei. We use the SD basis for the target (heavier cluster with mass number $A-a$) and the Jacobi basis for the projectile (mass number $a$). The translationally invariant kernels (\ref{H-kernel}) and (\ref{N-kernel}) can then be recovered by a transformation as described in Ref.~\cite{NCSM_RGM_long}. The use of the SD basis allows us to utilize all the shell-model techniques and codes.

In this contribution, we employ similarity-renormalization-group (SRG) ~\cite{SRG,Roth_SRG,Roth_PPNP} evolved chiral N$^3$LO $NN$ potentials~\cite{N3LO} (SRG-N$^3$LO) that are soft enough to allow us to reach convergence within about $12-16\hbar\Omega$ HO excitations in the basis expansion. The SRG evolution characterized by a parameter $\Lambda$ corresponds to a unitary transformation that transforms away off-diagonal $NN$ matrix elements in the momentum-space basis. The evolved $NN$ potential is phase-shift equivalent to the initial $NN$ potential. 

To guarantee the convergence of the NCSM/RGM calculation with respect to the HO basis expansion of both the cluster wave functions and the localized RGM integration kernels, we need to use a large $N_{\rm max}$ value. At present, we are able to perform calculations up to $N_{\rm max}=16$ for nuclei with $A<7$ (with $NN$ interactions). However, the use of large $N\hbar\Omega$ values is now feasible even for heavier nuclei thanks to the recent introduction of the importance truncated (IT) NCSM scheme~\cite{IT-NCSM,Roth09}. It turns out that many of the basis states used in the NCSM calculations are irrelevant for the description of any particular eigenstate, e.g., the ground state or a set of low-lying states. Therefore, if one were able to identify the important basis states beforehand, one could reduce the dimension of the matrix eigenvalue problem without losing predictive power. This can be done using an importance truncation scheme based on many-body perturbation theory~\cite{IT-NCSM}. 

\section{Applications}\label{results}

\subsection{Nucleon-$^4$He scattering}

\begin{figure}[b]
\begin{minipage}{7cm}
\includegraphics*[width=0.9\columnwidth]{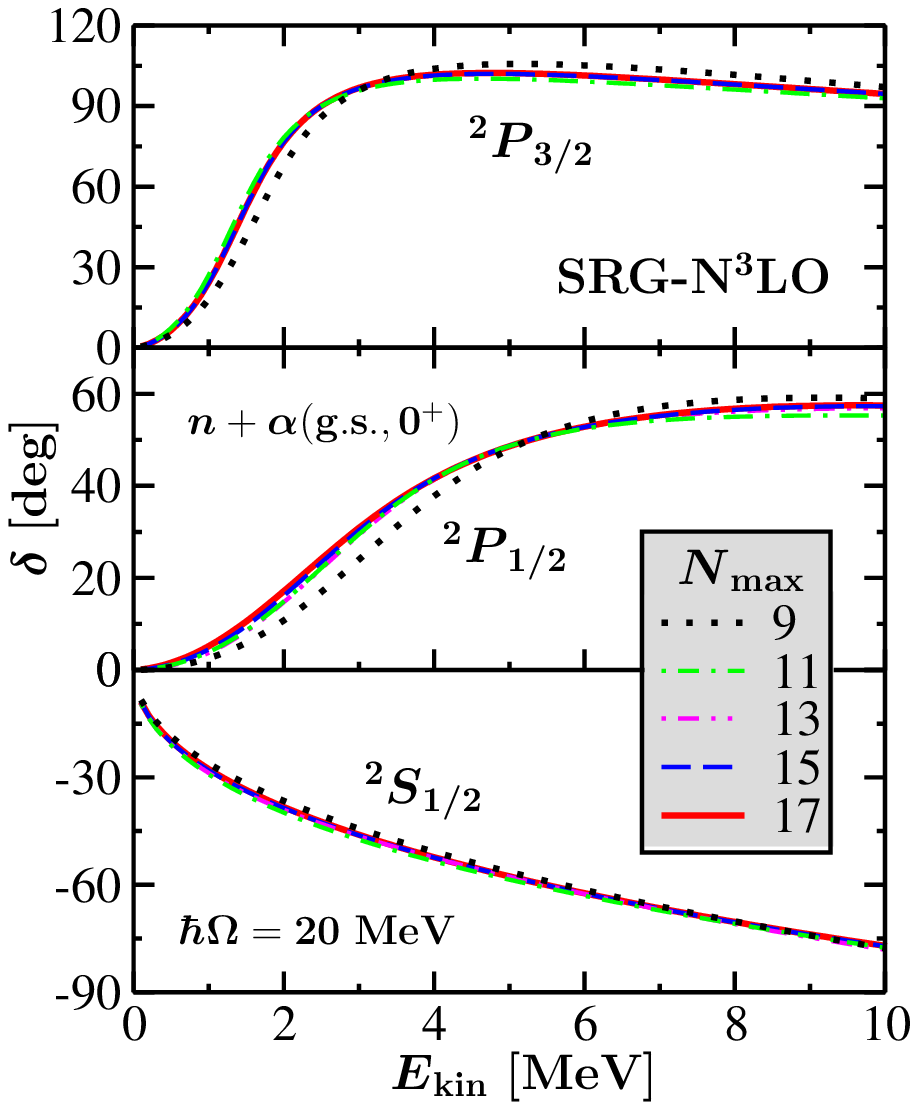}%
\end{minipage}
\begin{minipage}{7cm}
\includegraphics*[width=0.9\columnwidth, angle = -90]{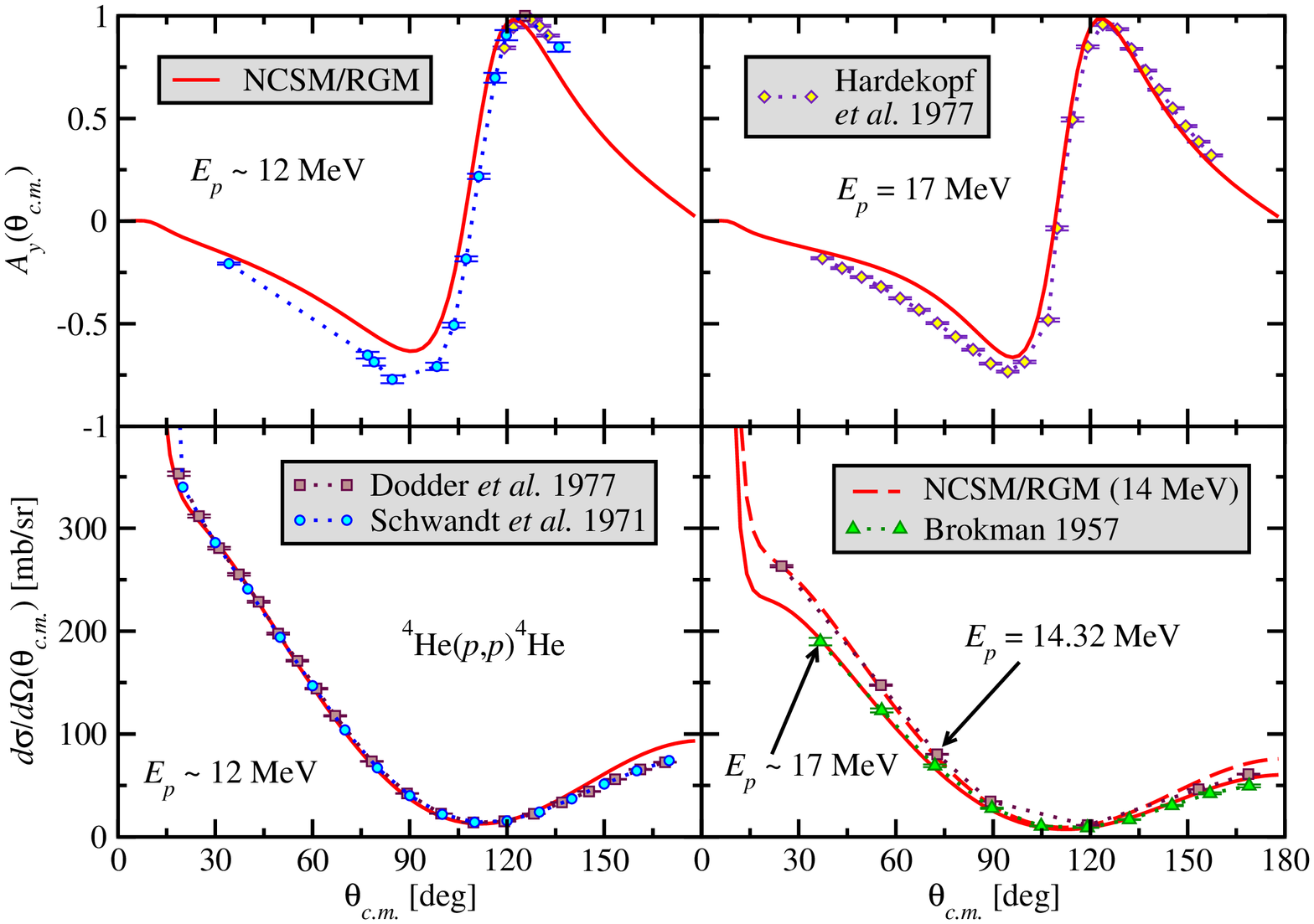}
\end{minipage}
\caption{Left figure: Dependence of the $n$-$^4$He phase shifts on the size of the HO basis expansion of the $^4$He wave functions and the localized parts of the integration kernels. Right figure: Calculated $p$-$^4$He differential cross section (bottom panels) and  analyzing power (top panels) for proton laboratory energies $E_p = 12, 14.32$ and 17 MeV compared to experimental data from Refs.~\protect\cite{Schwandt,Brokman,Dodder,Hardekopf}. See text for details of the calculations.}
\label{fig:nHe4_phaseconv}
\end{figure}
As the many-body wave function (\ref{trial}) is expanded in terms of binary-cluster eigenstates, the simplest system to investigate within the NCSM/RGM is the nucleon-$^4$He scattering. Limiting the expansion to just nucleon-$^4$He channel states with the $^4$He in its ground state (g.s.) is already an excellent approximation as the first excited state of $^4$He is at 20.21 MeV. A separate issue is the convergence with respect to the HO basis expansion. We performed extensive nucleon-$^4$He calculations with the SRG-N$^3$LO $NN$ potential with $\Lambda=2.02$ fm$^{-1}$ to check the convergence of our NCSM/RGM calculations. In the left part of Fig.~\ref{fig:nHe4_phaseconv}, we present  $n$-$^4$He phase-shift results for the $S$- and $P$-waves obtained using an HO basis expansion up to $N_{\rm max}=17$ for  for the localized parts of the NCSM/RGM integration kernels and for the $^4$He ground- and the first-excited $0^+ 0$ wave functions (since these states have positive parity, the $N_{\rm max}-1$ expansion is in fact used for the $^4$He eigenstates). As seen in the figure, the phase-shift convergence is excellent. In particular, the $N_{\rm max}=17$ and the $N_{\rm max}=15$ curves lie on top of each other.

The agreement of our calculated $n$-$^4$He and $p$-$^4$He phase shifts with the experimental ones is quite reasonable for the $S$-wave, $D$-wave and $^2P_{1/2}$-wave. The $^2P_{3/2}$ resonance is positioned at higher energy in the calculation and the corresponding phase shifts are underestimated with respect to the experimental results, although the disagreement becomes less and less pronounced starting at about 8 MeV. The observed difference is largely due to a reduction in spin-orbit strength caused by the neglect of the three-nucleon interaction in our calculations. More details are given in Ref.~\cite{NCSM_RGM_IT}, where we also benchmarked calculations with the full-space and the IT-space wave functions. As our calculated phase shifts agree with the experimental ones reasonably well above the center-of-mass energy of 8 MeV, we expect a similar behavior for cross section and analyzing power in that energy range. This is indeed the case as shown in the right part of Fig.~\ref{fig:nHe4_phaseconv}, where the NCSM/RGM $p-^4$He results are compared to various experimental data sets~\cite{Schwandt,Brokman,Dodder,Hardekopf} in the energy range $E_p \sim 12-17$ MeV.

\subsection{Neutron-$^3$H elastic cross section at 14 MeV}

\begin{figure}[b]
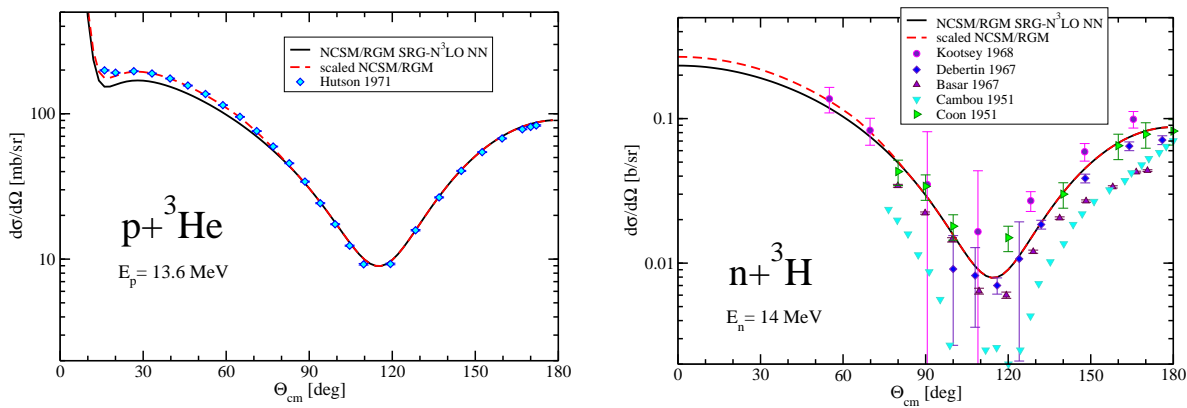

\begin{minipage}{7.5cm}
\includegraphics*[width=1.0\columnwidth]{dsigma_dOmega_pHe3_srg-n3lo0600_20_19_J0-4_NCSMonly_fig_13.6MeV.eps}
\end{minipage}
\hspace{1.pc}
\begin{minipage}{7.5cm}
\includegraphics*[width=1.0\columnwidth]{dsigma_dOmega_nH3_srg-n3lo0600_20_19_J0-4_14MeV_NCSMonly_fig.eps}
\end{minipage}
\caption{The $p$-$^3$He elastic differential cross section at 13.6 MeV (left panel) and the $n$-$^3$H elastic differential cross section at 14 MeV (right panel). The NCSM/RGM calculations (solid black line) described in the text are compared to the experimental data. The $p$-$^3$He data of Ref.~\protect\cite{Hutson71} are then used to scale the theoretical calculations at forward angles (dashed red line).}
\label{fig:n-3H}
\end{figure}
The fusion of deuterons with tritons in the $^3$H($d$,$n$)$^4$He reaction produces neutrons with laboratory energy $E_n$ of 14 MeV. When this reaction is studied at facilities like NIF, it is important to know precisely the cross sections of secondary reactions induced by the 14 MeV neutrons, like the elastic $n$-$^3$H cross section, in order to understand the measured neutron spectra. Experimental data for the elastic $n$-$^3$H reaction at this energy are not accurate. However, the elastic differential cross section for the mirror reaction, $p$-$^3$He, was measured with high accuracy at $E_p$=13.6 MeV~\cite{Hutson71}. We are in the position to apply NCSM/RGM calculations to relate the two cross sections. The energy of interest is above the triton and $^3$He breakup thresholds. Standard few-body calculations are difficult to apply in this case. As the NCSM/RGM approach uses an expansion in the cluster eigenstates there are no technical difficulties to use the approach above breakup threshold. Here, we limit ourselves to channel states with the triton and the $^3$He in their ground state. Obviously, the calculation is then only approximate, but the inaccuracy can be corrected by  comparing to the $p$-$^3$He data. This is shown in Fig.~\ref{fig:n-3H}. The solid black lines are the NCSM/RGM results obtained with the SRG-N$^3$LO $NN$ potential with $\Lambda=2.02$ fm$^{-1}$. The agreement with the experimental $p$-$^3$He differential cross section is very good at backward angles. At forward angles it is underestimated by up to 15\%. We can use the $p$-$^3$He data to deduce a smooth scaling factor and apply it to the calculated $n$-$^3$H differential cross section. The scaled cross sections are shown by dashed lines. The $n$-$^3$H differential cross section at $E_n$=14 MeV inferred in this way, agrees with the cross section obtained by means of an R-matrix analysis (also relying on the $p$-$^3$He data) within 10\%~\cite{Hale90,Hale_priv}. The integrated elastic cross section at 14 MeV obtained from the scaled NCSM/RGM calculation is 0.94 barn.

\subsection{Proton-$^7$Be scattering}

\begin{figure}[tb]
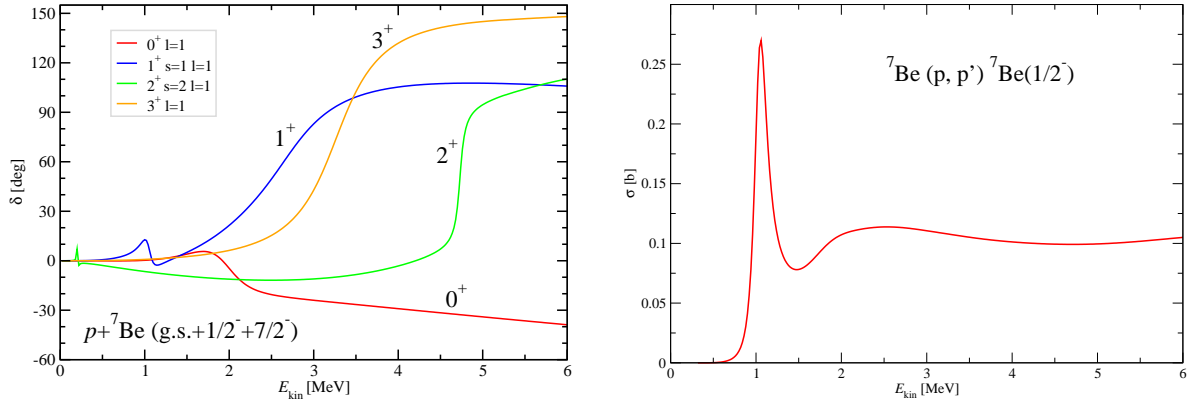

\begin{minipage}{7.5cm}
\includegraphics*[width=1.0\columnwidth]{phase_shift_pBe7_srg-n3lo2.0205_20_15_317_IT_Pwaves_fig.eps}
\end{minipage}
\hspace{1.pc}
\begin{minipage}{7.5cm}
\includegraphics*[width=1.0\columnwidth]{sigma_reac_pBe7_srg-n3lo2.0205_20_15_317_IT_gs-1m.eps}
\end{minipage}
\caption{$P$-wave diagonal phase shifts of the $p$-$^7$Be elastic scattering (left figure) and inelastic $^7$Be($p$,$p'$)$^7$Be(1/2$^-$) cross section (right figure). Energies are in the center of mass. For more details, see text.}
\label{fig:p-7Be}
\end{figure}
The $^7$Be($p$,$\gamma$)$^8$B capture reaction plays a very important role in nuclear astrophysics as it serves as an input for understanding the solar neutrino flux~\cite{Adelberger}. The S-factor extrapolation to astrophysically relevant energies depends, among other things, on the scattering lengths of the proton scattering on $^7$Be. Here, we report $p$-$^7$Be scattering calculations performed as a necessary preparatory step to eventually investigate the $^7$Be($p$,$\gamma$)$^8$B capture reaction. We performed NCSM/RGM calculations, described in detail in Ref.~\cite{NCSM_RGM_IT}, including $^7$Be ground state and $1/2^-$ and $7/2^-$ excited states,  using the SRG-N$^3$LO $NN$ potential with $\Lambda=2.02$ fm$^{-1}$. These latter wave functions were obtained by means of IT-NCSM calculations in the  $N_{\rm max}=12$ basis with an HO frequency of $\hbar\Omega=20$ MeV. The use of the IT-NCSM is essential here, as a full-space calculation can be performed only up to $N_{\rm max}=10$ which is not sufficient for reaching convergence. With this $NN$ potential, we do not find a bound state. As seen in the left part of Fig.~\ref{fig:p-7Be}, the lowest $2^+$ resonance, corresponding to the $^8$B ground state, lies at about 200 keV above threshold. In experiment, $^8$B is bound by 137 keV~\cite{TUNL_A8}. The calculated lowest $1^+$ resonance appears at about 1 MeV. It corresponds to the experimental $^8$B $1^+$ state at $E_x=0.77$ MeV (0.63 MeV above the $p$-$^7$Be threshold). This resonance dominates the inelastic cross section as seen in the right part of Fig.~\ref{fig:p-7Be}. We find $0^+$ and $2^+$ resonances not included in the recent $^8$B evaluation~\cite{TUNL_A8}. We note, however, that in the very recent Ref.~\cite{Mitchell10}, the authors claim the observation of low-lying $0^+$ and $2^+$ resonances, based on an R-matrix analysis of their $p$-$^7$Be scattering experiment. We also calculated the $S$-wave scattering length. In particular, we find $a_{02}=-10.2$ fm, close to the experimental value -7(3) fm from Ref.~\cite{Be7_scatl}. Our prospects for a realistic calculation of the $^7$Be($p$,$\gamma$)$^8$B capture are excellent. Here we found the $^8$B unbound by only 200 keV. We already tested that by repeating the described calculation with an $NN$ potential with a lower SRG parameter of $\Lambda=1.8$ fm$^{-1}$, the $2^+$ state of $^8$B becomes bound by 0.12 MeV, close to the experimental value. 

\subsection{Neutron-$^8$He scattering}

The neutron-rich helium isotopes beyond $^4$He provide an opportunity to investigate how nuclear structure develops as a function of increasing neutron number. The heaviest bound He isotope is $^8$He. The structure of the exotic $^9$He isotope is surrounded by controversy. Its ground state was considered to be a narrow resonance at about $1.3$ MeV above the $n$-$^8$He threshold~\cite{Bohlen99}. A two-proton knockout reaction experiment~\cite{Chen01} gave evidence that the ground state of $^9$He is an $S$-state around 0.2 MeV above the $n$-$^8$He threshold with an upper limit for the scattering length of $a_0<-10$ fm. Recent studies of the two-proton removal from $^{11}$Be and $^{12}$C~\cite{AlFalou07} resulted in a value for $a_0$ rather in the range of $-2$ to $0$ fm. Similarly, the analysis of proton-knockout reactions from $^{11}$Li found the $S$-wave scattering length of $a_0\approx -3$ fm~\cite{Johansson10}. Such a small absolute value of the scattering length  suggests that the $S$-wave state cannot be classified as a resonance. 

\begin{figure}[tb]
\begin{minipage}{7.5cm}
\includegraphics*[width=1.0\columnwidth]{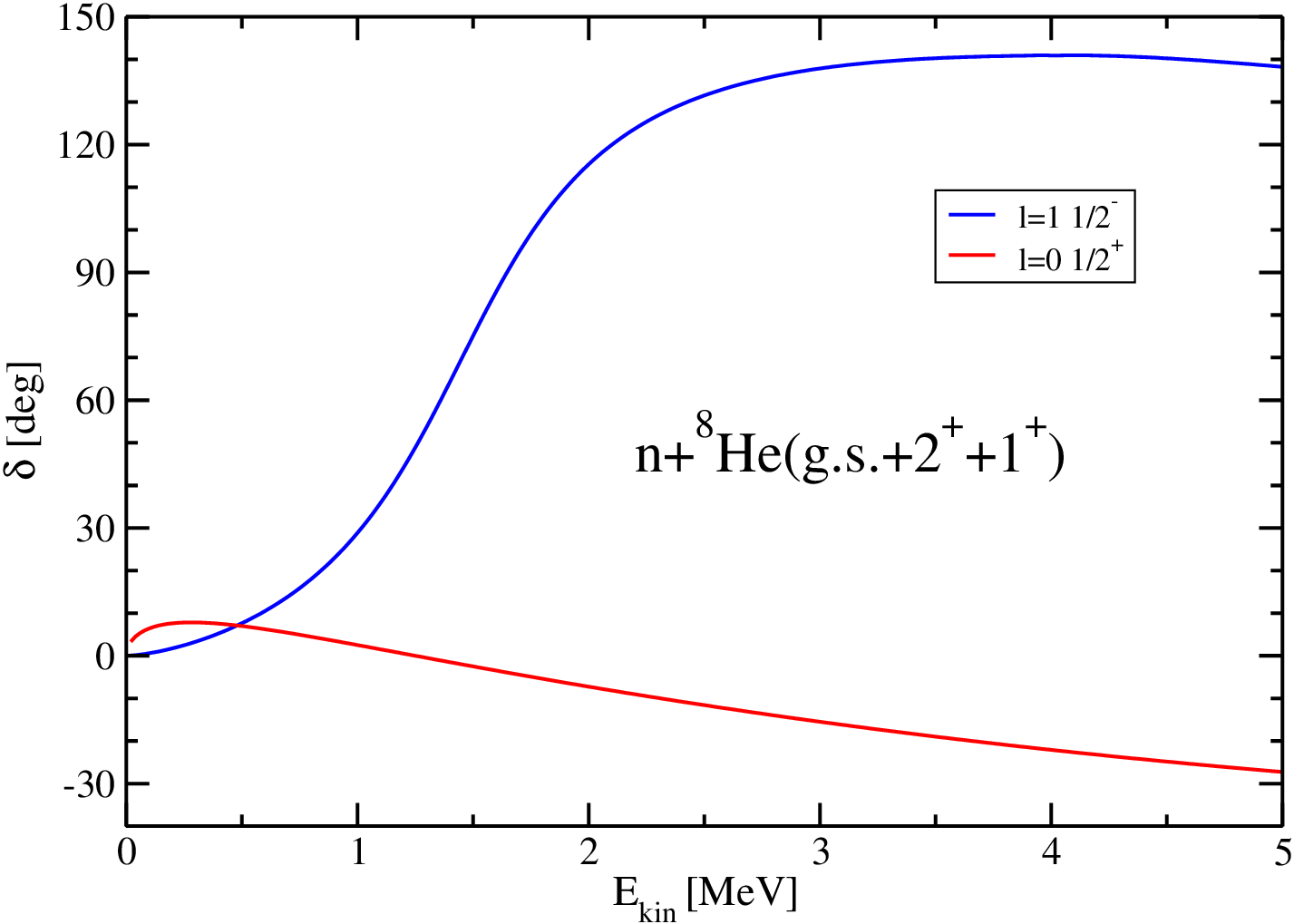}
\caption{$S$-wave and $P$-wave diagonal phase shifts of the $n$-$^8$He elastic scattering. The energy is in the center of mass.  For more details, see text.}
\label{fig:n-He8}
\end{minipage}
\hspace{1.pc}
\begin{minipage}{7.5cm}
\includegraphics*[width=1.0\columnwidth]{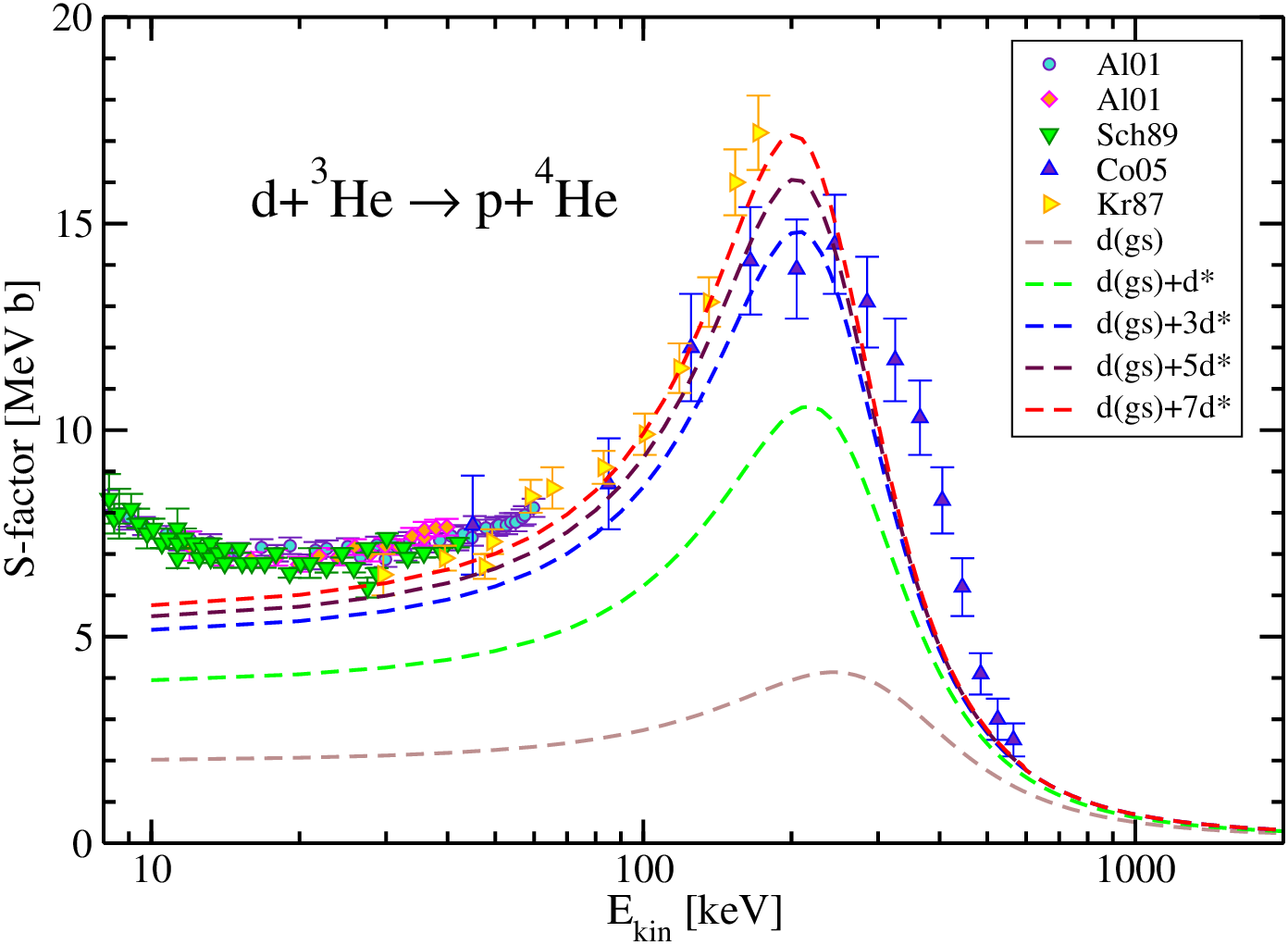}
\caption{Calculated S-factor of the $^3$He($d$,$p$)$^4$He fusion reaction compared to experimental data. Dependence on the number of included $^3S_1-^3D_1$ channel pseudo states. For more details, see text.}
\label{fig:d-He3}
\end{minipage}
\end{figure}

To confront this situation with a microscopic theory, we performed NCSM/RGM calculations that included the $^8$He ground state and the $2^+$ and $1^+$ excited states using the SRG-N$^3$LO $NN$ potential with $\Lambda=2.02$ fm$^{-1}$. The  $N_{\rm max}=8$ basis and the HO frequency of $\hbar\Omega=16$ MeV were employed. Our calculated $S$-wave and $P$-wave phase shifts are presented in Fig.~\ref{fig:n-He8}. We find a resonance in the $1/2^-$ $P$-wave at about 1.6 MeV in reasonable agreement with the experimental value of about 1.3 MeV~\cite{Bohlen99}. In the $1/2^+$ $S$-wave we see a weak attraction rather than a resonance. Our calculated scattering length, $a_0\sim -2$ fm, is close to the experimental values from Refs.~\cite{AlFalou07} and \cite{Johansson10}. In summary, our  present results favor a $1/2^-$ ground state, although more $n$-$^8$He calculations are under way. 

\subsection{Deuteron-$^3$H and deuteron-$^3$He fusion}

The deuterium-tritium reaction is important for possible future fusion energy generation. It is used in laser-induced fusion at NIF and magnetic-confinement fusion at ITER. Even though it has been well studied experimentally, its first principles theoretical understanding is important. The $^3$H($d$,$n$)$^4$He and its mirror reaction $^3$He($d$,$p$)$^4$He are also of interest for understanding primordial nucleosynthesis. In addition, the  $^3$He($d$,$p$)$^4$He is one of the few reactions to present strong electron screening effects. 

Here we present the first {\it ab initio} calculations for these reactions. We have further developed the NCSM/RGM formalism to include the deuteron projectile (i.e. deuteron-nucleus channels) and the coupling between the single-nucleon-projectile and the deuteron-projectile channels. To calculate the integration kernels~(\ref{H-kernel}) and (\ref{N-kernel}) in the SD formalism, it is necessary to calculate up to the three-body density of the target nuclei (here $^4$He, $^3$H and $^3$He). Our first results, that are still preliminary, were obtained with the SRG $NN$ interaction with $\Lambda=1.5$ fm$^{-1}$ in the HO model spaces up to $N_{\rm max}=13$ and HO frequency of $\hbar\Omega=14$ MeV.

The astrophysical S-factor for the $^3$He($d$,$p$)$^4$He reaction from beam-target experiments is compared to {\it ab initio} NCSM/RGM calculations for bare nuclei in Fig.~\ref{fig:d-He3}. When only the ground states of the involved nuclei are considered, the cross section is severely underestimated. Because it is so weakly bound, the polarization and virtual breakup of the deuteron cannot be neglected. As seen in the figure, the calculated S-factor improves dramatically with the inclusion of the virtual breakup of the deuterium, obtained by means of excited deuterium pseudo-states ($d^*$).  We observe a slightly different shape of the peak than that suggested by the ``Trojan-horse'' data from Ref.~\cite{Co05}. Also, no low-energy enhancement is present in the theoretical results contrary to the beam-target data of~\cite{Al01} affected by the electron screening. Our $^3$H($d$,$n$)$^4$He results, that will be published in a forthcoming paper, show a still stronger sensitivity to the deuteron polarization due to the lower position of the resonance peak.

\section{Conclusions}\label{concl}

Great progress has been made in the development of microscopic and {\it ab initio} approaches to nuclear scattering, reactions and the description of weakly bound states. The accuracy of few-body methods improved and their ability to treat non-local and three-nucleon interactions has been extended. There are promising developments in methods applicable to systems of more than four nucleons. Continuum effects can now even be investigated in semi-magic nuclei beyond the $p$-shell. 

We focused on recent and new results obtained within the NCSM/RGM approach. We discussed results of nucleon scattering on various light nuclei as well as the latest extension to include the deuteron projectile and the coupling of deuteron-projectile and nucleon-projectile channels. Our results are very promising and prospects are good to extend the approach by including three-nucleon forces (both genuine and those induced by the SRG evolution~\cite{Jurgenson09}), three-nucleon projectiles, and eventually the three-body continuum.

\ack
Prepared in part by LLNL under Contract DE-AC52-07NA27344.
Support from the U.\ S.\ DOE/SC/NP (Work Proposal No.\ SCW0498), LLNL LDRD grant PLS-09-ERD-020, and from the U.\ S.\ Department of Energy Grant DE-FC02-07ER41457 is acknowledged. This work is supported in part by the Deutsche Forschungsgemeinschaft through contract SFB 634 and by the Helmholtz International Center for FAIR within the framework of the LOEWE program launched by the State of Hesse.

\end{document}